\documentclass[10pt]{iopart}

\usepackage{iopams} 
\usepackage{graphicx}
\usepackage{xcolor}
\begin{document}

\title[]{\Large On the impact of temperature gradient flattening and system size on heat transport in microtearing turbulence}

\author{Ajay C. J.$^1$, Ben McMillan$^1$ and M. J. Pueschel$^{2,3}$}

\address{$^1$Centre for Fusion, Space and Astrophysics, Department of Physics, University of Warwick, CV4 7AL, Coventry, UK\\
$^2$Dutch Institute for Fundamental Energy Research, 5612 AJ Eindhoven, The Netherlands\\
$^3$Eindhoven University of Technology, 5600 MB Eindhoven, The Netherlands}
\ead{ajay.chandrarajan-jayalekshmi@warwick.ac.uk}
\vspace{10pt}
\begin{indented}
\item[]March 2023
\end{indented}

\begin{abstract}
Microtearing instability is one of the major sources of turbulent transport in high-$\beta$ tokamaks. These modes lead to very localized transport at low-order rational magnetic field lines, and we show that flattening of the local electron temperature gradient at these rational surfaces plays an important role in setting the saturated flux level in microtearing turbulence. This process depends crucially on the density of rational surfaces, and thus the system-size, and gives rise to a worse-than-gyro-Bohm transport scaling for system-sizes typical of existing tokamaks and simulations. 
\end{abstract}

%
%
%
%
\ioptwocol

\section{Introduction}
Confinement in tokamaks is enabled by magnetic field lines that trace out nested toroidal surfaces. On {\it rational} surfaces, the field lines connect back to themselves after integer numbers of poloidal and toroidal turns; certain electromagnetic plasma instabilities, such as the microtearing modes of interest here, are localized near these rational surfaces, and break the nested topology by forming magnetic islands~\cite{Hazeltine1975,Drake1980,Connor1990}. 
Microtearing modes may  have a significant impact on confinement, especially in high-$\beta$ spherical tokamaks~\cite{Gryaznevich1998} where electromagnetic instabilities tend to be more unstable~\cite{Applegate2004,Guttenfelder2011,Patel2021}, and hence understanding microtearing transport is crucial for designing large spherical tokamak reactors such as STEP~\cite{Wilson2020}. They also are believed to be important in the pedestal in conventional tokamaks~\cite{Hatch2021}.


Microtearing modes are characterized by radially-narrow parallel electron current layers that are driven resonantly at the rational surface and associated magnetic islands. While various branches of microtearing modes have been identified, including those driven by the time-dependent thermal force~\cite{Drake1981} or by curvature~\cite{Carmody2013,Hamed2018_2}, and are present in various collisionality regimes~\cite{Drake1977,Gladd1980,Predebon2013,Swamy2015}, the electron temperature gradient remains a necessary condition for instability in all cases. Previous works have reported various microtearing saturation mechanisms: via energy transfer to long wavelengths~\cite{Drake1980} or to short wavelengths~\cite{Doerk2011}, via cross-scale coupling with electron temperature gradient modes~\cite{Maeyama2017_2}, by background shear flow~\cite{Guttenfelder2011} or zonal fields~\cite{Pueschel2020}, etc. However, despite these advances, predicting saturation levels remains a challenging task. 
We investigate a previously studied~\cite{Doerk2011,DoerkPhD,Xie2020} microtearing turbulence case (known to be numerically tractable) in simple geometry, resolving only the ion scales, in the absence of background flow shear, and demonstrate a saturation mechanism where the magnetic islands associated with the resonant current layers flatten the electron temperature gradient, thereby reducing the linear drive at the rational surfaces.

The radial width of the resonant region at the rational surfaces is generally of the order of few ion Larmor radii and is set by the parallel correlation length in linear theory which scales with the square root of the mass ratio between ions and electrons~\cite{Hardman2022}. Nonlinearly, the flux associated with these modes, although slightly broadened, is still localized near the rational surfaces. This is already known to be important in setting global flux levels in the pedestal~\cite{Hatch2021}. In the most extreme case, if turbulent diffusivity is sufficiently large and localized near low-order rational surfaces, the system will remove the local driving gradients, increase the gradients away from the low-order rational surfaces, and saturate in a zero-flux state. In this work, we find a less extreme version of this process occurring in a standard microtearing regime. We make a scaling argument to quantify this effect and suggest that future reactor-size devices subject to microtearing turbulence may perform worse that expected.

In simulations of microtearing turbulence, {\it runaway} states quite often develop with extreme flux levels. In this work, we do not directly address how and where this process occurs, although we see this occur in certain of our simulations.

We proceed by demonstrating the strong electron-temperature-gradient flattening at low-order rationals in gyrokinetic simulations, and showing that this allows saturation by reducing mode drive. We test the impact of the various saturation mechanisms by suppressing zonal modulations and show the dominance of the temperature corrugations. Lastly, we consider system-size scaling and explain the origin of a non-gyro-Bohm scaling.

\section{Simulation set-up.}
Our numerical investigation uses \emph{flux-tube} and \emph{global} \textsc{Gene} gyrokinetic simulations~\cite{GENE1,GENE2}. 
In the flux-tube version~\cite{Beer1995}, the background quantities and their gradients are assumed constant over its radial domain~\cite{Scott1998}. The global simulations, on the other hand, can accommodate more realistic background profile variations, but in general may be more computationally expensive, and furthermore, setting appropriate boundary conditions, sources etc. can prove difficult~\cite{GENE2,GoerlerPhD}. By comparing these two simulation types, we are able to determine whether the more complicated global physics is playing an important role~\cite{McMillan2010}.

The simulations use a field-aligned coordinate system~\cite{Beer1995} where $x$ is the radial coordinate, $y$ the binormal coordinate and $z$ the parallel coordinate. The binormal wavenumber $k_y=nq_0/x_0$, where $n$ is the toroidal mode number and $q_0$ is the safety factor at the radial position $x_0$ where the simulation is centered. Parallel velocity $v_\parallel$ and magnetic moment $\mu$ are the velocity space coordinates. 
We use the simple microtearing-dominated equilibrium used in Refs.~\cite{Doerk2011,Xie2020}, modelling the outer core region of a tokamak. Concentric circular flux-surface geometry \cite{Lapillonne2009} is considered with an inverse aspect ratio $\epsilon=x_0/R=0.15$. An electron-ion mass ratio of $m_i/m_e=1836$, temperature ratio of $T_{i,0}/T_{e,0}=1$ and normalized electron pressure of $\beta_e=0.4\%$ are considered. To model collisions, the linearised Landau operator is used with an electron-ion collision frequency $\nu_{ei}/(v_{th,e}/R)=0.02$, where $v_{th,s}=(T_{s,0}/m_s)^{1/2}$ is the thermal velocity of species $s$. $\delta B_\parallel$ fluctuations are not included.

In the \emph{flux-tube} simulations, a safety factor of $q_0=3$ and magnetic shear $\hat{s}=1$ are considered. The inverse of the density, ion temperature and electron temperature background gradient scale lengths, normalized to the major radius $R$, are $R/L_n=1$, $R/L_{T_i}=0$ and $R/L_{T_e}=4.5$, respectively. The standard nonlinear flux-tube simulation considered in this work has a minimum binormal wavenumber of $k_{y,\min}\rho_i=0.02$ ($L_y=314.2\rho_i$), radial width of $L_x=150\rho_i$ and grid resolutions given by $N_x\times N_{y}\times N_z\times N_{v_\parallel}\times N_{\mu} = 192\times 48\times 16\times 36 \times 8$. It is run until normalized time $2000 R/v_{th,i}$ (max lin. growth rate $= 0.018 v_{th,i}/R$) and saturates to give a gyro-Bohm normalized electron electromagnetic heat flux of $Q_{e,em}/Q_{GB}=7.9$ (see Fig.~\ref{Fig5}). 

The corresponding \emph{global} simulations, centered at $x_0=0.5a$, span a radial width of either $L_x=0.15 a$ or $0.3 a$, where $a$ is the tokamak minor radius. A quadratic  q-profile of the form $q(x)=1.5+6(x/a)^2$ is considered. The radial background temperature and density profiles are of the form $A_s={\rm exp}[- \kappa_{As} \ \epsilon\  \Delta A_s\ {\rm tanh}((x-x_0)/(a\Delta A_s))]$ where $A_s$ represents the temperature or density of species $s$; $\kappa_{n}=1$, $\kappa_{Ti}=0$, $\kappa_{Te}=4.5$ and $\Delta n=\Delta T_i=\Delta T_e=0.3$. The numerical resolutions for the simulation with $\rho^\star=\rho_i/a=0.004$ and $L_x=0.3a$ are $N_x\times N_{y}\times N_z\times N_{v_\parallel}\times N_{\mu} = 128\times 36\times 16\times 36 \times 16$. Krook heat and particle sources (see Ref.~\cite{Lapillonne2011_2} for details) are also employed with a source rate of $\gamma_h=\gamma_p=0.015 v_{th,i}/R$, and with radial smoothing over $0.09a$ applied so these operators maintain the global-scale profiles, but do not damp finer-scale corrugations.
The result is a radially smooth heating profile, and thus a radially smooth quasi-steady state heat flux, as would be expected in experiment, and consistent with local simulations. Doubling or halving this source rate is found to have little effect on the time-averaged density and temperature profiles, and the heat flux-levels change only by $20\%$ at most. This is unlike the global simulations of microtearing of \cite{Hatch2021} that had heat fluxes that were sensitive to the source level, and had very radially peaked heat fluxes near low-order rationals.

\section{$T_e$ flattening at low-order rational surfaces.}
\begin{figure}[t]     	
\includegraphics[width=\columnwidth,trim={5ex 23ex 0 55ex},clip]{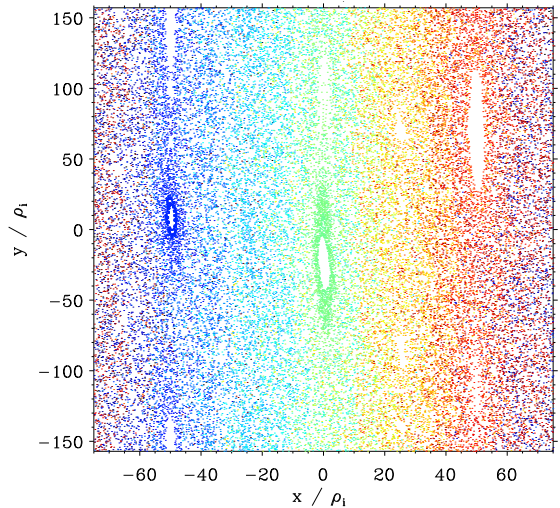}
\caption{Poincar{\'e} plot of magnetic field lines intersecting the outboard midplane for the standard nonlinear simulation at $t v_{th,i}/R=1750$.}
\label{Fig2}
\end{figure}

Modes at a specific toroidal mode number $k_y$ create magnetic islands around the resonantly driven current layers at their respective mode rational surfaces (MRSs). Note that the distance between MRSs for a given $k_y$ is $1/({\hat{s}k_y})$ in flux-tube simulations. The MRS of all $k_y$ radially align at the lowest-order mode rational surface (LMRS), where the magnetic islands can persist even in the turbulent phase. For the standard nonlinear flux-tube simulation, this can be seen at the LMRSs at $x/\rho_i=-50,\ 0$ and 50 in the Poincar{\'e} plot in Fig.~\ref{Fig2}. The Poincar{\'e} plot records the positions where each magnetic field line crosses the outboard midplane on successive poloidal turns \cite{Nevins2011,Pueschel2013_2}. Each color denotes an individual field line. Away from the LMRSs, the MRSs of each $k_y$ are radially misaligned and the overlapping magnetic islands give rise to ergodic regions.

\begin{figure}[b]     		
\includegraphics[scale=0.5]{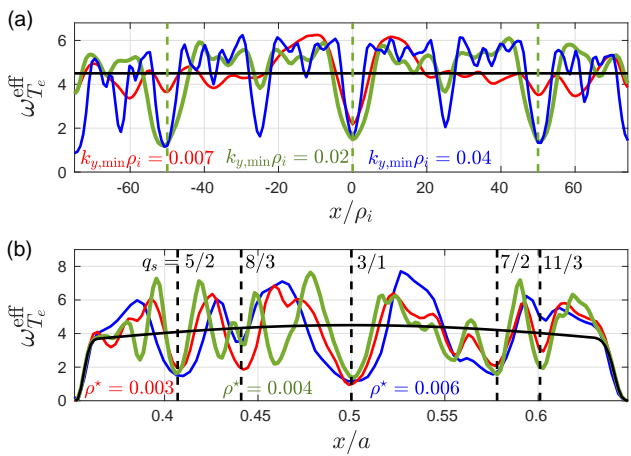}
\caption{
Time-averaged $\omega^{\rm eff}_{T_e}$ as a function of the radial coordinate. (a) Flux-tube simulation scan over $k_{y,\min}\rho_i$. Dashed lines denote position of LMRSs for the $k_{y,\min}\rho_i=0.02$ run. (b) Global simulation scan over $\rho^\star$. Dashed lines denote specific $q_s=m/n$ rational surfaces. Solid black line denotes the background gradient.}
\label{Fig3}
\end{figure}

As the electrons move swiftly along the parallel direction following the perturbed magnetic field associated with the islands at the low-order MRSs, they also undergo periodic radial excursions. This leads to a short-circuit of the perturbed $T_e$ profile, leading to its flattening. This can be seen in Fig.~\ref{Fig3}(a), where the green curve denoting the time-averaged effective temperature gradient $\omega^{\rm eff}_{T_e}$ is plotted as a function of the radial coordinate for the standard nonlinear simulation. $\omega^{\rm eff}_{T_e}$ is defined as the sum of the contributions from the background temperature gradient and the time-averaged zonal perturbed temperature gradient, \emph{i.e.}
\begin{equation}
\omega^{\rm eff}_{T_e}=-\frac{dT_{0,e}/dx}{T_{0,e}/R}-\frac{\langle\partial \delta T_{e}/\partial x\rangle_{yzt}}{ T_{0,e}/R}.
\end{equation}
The perturbed temperature is defined as $\delta T_e=(m_e/n_0)\int v^2 \delta f_e d^3v - (T_{e,0}/n_0) \int \delta f_{e} d^3v$, where $\delta f_e$ is the perturbed electron distribution function, and the flux-surface average, denoted by $\langle\cdot\rangle_{yz}$, extracts the zonal part. 

Time-averaged $\omega^{\rm eff}_{T_e}$ in global simulations (both $L_x=0.15a$ and $0.3a$) too show similar flattenings at low-order rational surfaces, as shown in Fig.~\ref{Fig3}(b) for the runs with $L_x=0.3a$. Note that for the two low-$\rho^\star$ global simulations having $L_x=0.3a$, the time-average is taken over the initial saturated state before they `run away'; more on this in Sec~\ref{SecSystemSize}.

One may also understand the temperature flattening as a consequence of turbulence self-interaction - a mechanism where modes that are significantly extended along the field line `bite their tails' at the rational surfaces~\cite{Justin2020,AjayCJ2020}. In the case of microtearing modes, the parallel electron heat current density $q_{e,\parallel}=\int v_\parallel^3\delta f_{e}d^3v$ which is extended along the field line, interacts with the $A_\parallel$ of the same eigenmode, to drive zonal parallel electron temperature perturbations $\langle \delta T_{e,\parallel}\rangle_{yz}$, leading to its flattening at MRSs. Here, $\delta T_{e,\parallel}$ is defined as $\delta T_{e,\parallel}=(m_e/n_0)\int v_\parallel^2 \delta f_{e} d^3v$ for convenience. 

\begin{figure}[t]
\centering
\includegraphics[width=\columnwidth]{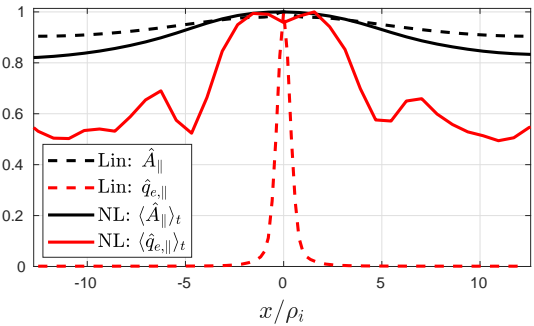}
\caption{Parallel vector potential $\hat{A}_{\parallel,k_y}$ (black) and parallel electron heat current $\hat{q}_{e,k_y}$ (red) for the microtearing mode with $k_y\rho_i=0.04$. The linear eigennmode is shown in dashed lines and the time-averaged nonlinear data in solid lines, all normalized by their maximum.}
\label{Figqpar}
\end{figure}

Taking the $v_\parallel^2$ moment and the flux-surface average of the gyrokinetic Vlasov equation, one arrives at an equation for the time evolution of the zonal $\delta T_{e,\parallel}$. Since we are interested in the microtearing mode, only the electromagnetic ($\propto A_\parallel$) nonlinear term need to be considered. Furthermore, the gyro-average over $A_\parallel$ may be ignored given that it has a radially broad structure as shown in Fig.~\ref{Figqpar}. These approximations helps one to obtain the relation
		\begin{equation}
		\frac{\partial \langle \delta T_{e,\parallel}\rangle_{y}}{\partial t} 
		\approx -\frac{m_e}{n_0}\frac{1}{\mathcal{C}}
		\frac{\partial}{\partial x}
		\sum_{k_y}ik_y\hat{q}_{e,\parallel,k_y}\hat{A}^*_{\parallel,k_y},
		\end{equation}
where the constant $\mathcal{C}=B_0/|\nabla x\times \nabla y|$. The linear structures of $\hat{q}_{e,\parallel,k_y}$ and $\hat{A}_{\parallel,k_y}$ for $k_y\rho_i=0.04$ are plotted with dashed lines in Fig.~\ref{Figqpar}. The product of the two, proportional to a linear heat flux contribution, drives a zonal $\delta T_{e,\parallel}$ that leads to the flattening of the parallel electron temperature at each MRS. The same process repeats for the perpendicular electron temperature.
		
However, note a significant broadening of the time-averaged $\hat{q}_{e,\parallel,k_y}$ in nonlinear simulation, also shown in Fig.~\ref{Figqpar} with a solid red line. A detailed description of this nonlinear broadening mechanism is given in Refs.~\cite{AjayCJ2020,AjayCJ2021} and can be summarized as follows. The radially narrow linear eigenmode structures lead to extended tails in $k_x-$Fourier space and in ballooning representation (called `giant tails' \cite{Hallatscheck2005}). However, in a nonlinear simulation, only the first few linearly coupled $k_x$-Fourier modes starting from $k_x=0$ of the eigenmode are able to retain their linear characteristics, \emph{i.e.}, their high amplitudes and relative phase differences with the $k_x=0$ mode, whereas the Fourier modes further away in the tail undergo a significant reduction in their amplitudes as a result of nonlinear couplings, implying a broadening in real space. The width of the flattened electron temperature is therefore also broadened.

\section{Microtearing stability with corrugated background gradients}
Now, we consider the linear stability of microtearing modes when the effective electron temperature gradient $\omega^{\rm eff}_{T_e}$ has local flattenings at LMRSs. This is equivalent to the tertiary instability analysis of zonal flows~\cite{Pueschel2013_3}, except with a fixed temperature corrugation rather than a zonal flow pattern.

\begin{figure}[t]
\centering
\includegraphics[scale=0.48]{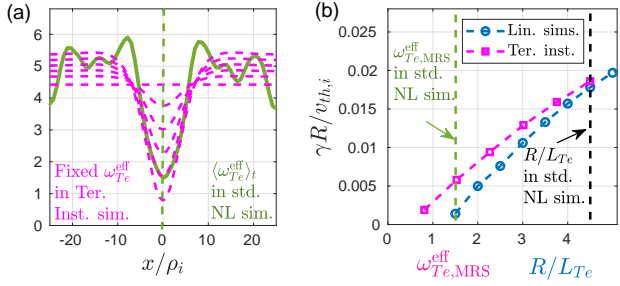}
\caption{(a) Magenta: Various fixed $\omega^{\rm eff}_{T_e}$ considered for the tertiary instability simulations. Solid green: Time-averaged $\omega^{\rm eff}_{T_e}$ in the standard nonlinear simulation. Dashed green: Position of the MRS. (b) Magenta: Growth rate in tertiary instability simulations as a function of $\omega^{\rm eff}_{T_e,{\rm MRS}}$. Blue: Growth rate in linear simulations as a function of $R/L_{Te}$.}
\label{FigZonInit}
\end{figure}

We perform the same nonlinear local simulation but with only two $k_y$ - the zonal $k_y=0$ mode and an unstable microtearing mode $k_y=0.02\rho_i^{-1}$. The zonal mode is reinitialised at each time-step such that the effective gradient $\omega^{\rm eff}_{T_e}$ remains constant in time and has local flattenings at LMRSs as shown by the dashed magenta curves in Fig.~\ref{FigZonInit}(a), and resembling the time-averaged $\omega^{\rm eff}_{T_e}$ in the standard nonlinear simulation shown by the green curve. The $k_y\rho_i=0.02$ mode is then initialised to a low seed-level amplitude and let to evolve in time atop the background of the fixed effective gradient. The simulation is then repeated for different values of $\omega^{\rm eff}_{T_e}$ at the MRS, and the growth rate of the microtearing mode is measured for each case.
	
Since the resonant current drive leading to the microtearing instability is also localized at the MRS, we expect the growth rate of the microtearing modes considered in these tertiary instability simulations to be set mostly by the effective gradient $\omega^{\rm eff}_{T_e,\rm MRS}$ at the MRS, \emph{i.e.}, the temperature gradient away from MRS is of little significance. This is verified in Fig.~\ref{FigZonInit}(b) by the close match between the growth rate obtained from the tertiary instability simulations plotted as a function of $\omega^{\rm eff}_{T_e,\rm MRS}$ (magenta) and the growth rate obtained from standard linear simulations plotted as a function of $R/L_{T_e}$ (blue). The figure also suggests that the time averaged $\omega^{\rm eff}_{T_e,\rm MRS}$ at the LMRSs in standard nonlinear simulation is set by the critical gradient of the instability. For $k_y>k_{y,\min}$, while the modes at LMRSs are made almost fully stable, those at MRSs away from the LMRSs, with lesser flattenings, are made less stable. In general, by reducing the local drive of microtearing modes at the rational surfaces, the system saturates to a state with lesser flux.
	 
\section{Removing zonal modulations}
To further investigate the role of electron temperature flattening on saturation, a nonlinear flux-tube simulation is run while eliminating any local modifications to the temperature gradient. This is achieved by redefining the zonal component of the electron distribution function as $\left\langle\delta f_{e}^{\rm mod}\right\rangle_{yz}=\left\langle\delta f_e\right\rangle_{yz} - K(x)\left[m_ev^2/2T_{e,0} - 1.5\right]\left\langle f_{M,e}\right\rangle_{yz}$, where $f_{M,e}$ is the electron background distribution function - a homogeneous local Maxwellian. Note that $K(x)$ is only a function of $x$ and is set at each time-step such that $\langle\delta T_{e}\rangle_{yz}=0$, and therefore $\omega^{\rm eff}_{T_e}=R/L_{Te}$ throughout the simulation. The heat flux $Q_{e,em}/Q_{GB}=28.1$ in this simulation is many times higher than $Q_{e,em}/Q_{GB}=7.9$ in the original standard nonlinear simulation, as shown in Fig.~\ref{Fig5}, confirming that electron temperature flattening indeed plays a significant role in turbulence suppression.
    
Another way to reduce the electron temperature flattening is by weakening the self-interaction process by increasing the parallel length $L_z=2\pi N_{pol}$ of the simulation volume~\cite{Justin2020,AjayCJ2020,Ishizawa2014}, where $N_{pol}$ indicates the number of times the flux-tube wraps around poloidally before connecting back to itself. Given that self-interaction essentially results from `modes biting their tails', to correctly capture its effects, the domain length along a field line at a rational surface must be correctly captured by modelling the full flux-surface, and hence $N_{pol}$ and the minimum toroidal mode number $n_{\min}$ both need to be set to 1~\cite{Justin2020}. Therefore, by increasing $N_{pol}$, we are unphysically weakening the self-interaction process and the resulting electron temperature flattening. Doubling $N_{pol}$ (which also doubles the denisty of LMRSs) is found to weaken the temperature flattening from $\approx 70\%$ to $\approx 20\%$ at LMRSs, leading to an increase in the flux level as shown in Fig.~\ref{Fig5}.

While these results confirm that the local flattening of electron temperature is crucial for correctly predicting the saturated turbulent state, the fact that these simulations, either with fully eliminated or weakened electron temperature flattenings, did saturate, indicates the presence of other, less dominant, saturation mechanism(s). Deleting the zonal electrostatic potential $\Phi$ or the zonal $A_\parallel$ in simulations changes the flux levels at most by 12\%, implying that zonal flows and fields do not play a significant role in saturation in the case considered. For similar parameters, Ref.~\cite{Doerk2011} shows that the free-energy flow to short wavelengths could be another saturation mechanism.

\section{Effect of system-size}
\label{SecSystemSize}
Given that microtearing turbulence saturation via temperature flattening happens primarily at rational surfaces, the separation distance between them and the corresponding finite system-size effect is crucial. To study this in detail, we first perform two system-size scans using global simulations, choosing radial width $L_x=0.15a$ for one scan and $L_x=0.3a$ for the other. 
As mentioned in the previous section, to correctly capture the effects of self-interaction including temperature flattening in these simulations, the minimum toroidal mode number $n_{\min}$ is set to 1. 

\begin{figure}[t]
\centering
\includegraphics[width=\columnwidth]{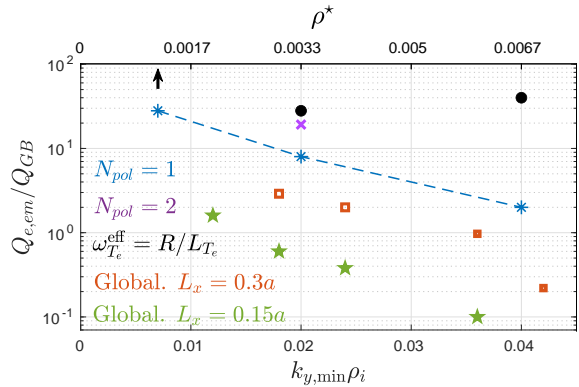}
\caption{Blue asterisks and violet cross denote the time-averaged gyro-Bohm normalised electron electromagnetic heat flux $Q_{e,em}/Q_{GB}$ as a function of $k_{y,\min}\rho_i$ in flux-tube simulations with $N_{pol}=1$ and $N_{pol}=2$ respectively. Top axis indicates $\rho^\star$ considering $r_0/a=0.5$. Black markers indicate flux-tube simulations with $\omega^{\rm eff}_{T_e}=R/L_{T_e}$. The arrow indicates that the simulation saturates only transiently. Orange squares and green stars denote global simulations with $L_x=0.3a$ and $L_x=0.15a$ respectively. Open squares indicate a runaway at a later stage.}
\label{Fig5}
\end{figure}

In general, the heat flux increases with increasing system-size as shown in Fig.~\ref{Fig5}. In the later part of this section, we explain why this is consistent with a saturation mechanism that depends on zonal temperature flattening at MRSs. Note that two sets of global simulations have different flux levels, and both are also below the local simulation flux levels. This difference can't be explained in terms of the zonal-temperature-gradient flattening mechanism explored here, so suggests there is an additional system-size effect dependent on how the radial domain is treated.

For the global simulation with $L_x=0.3a$ at the two larger system sizes denoted by open squares, the fluxes saturate to a turbulent steady state for at least a duration of $400R/v_{th,i}$, after which they undergo a `runaway' similar to what has been reported in Ref.~\cite{Hatch2021}. 
The runaway persists even when the grid resolutions are increased and the time-step is decreased, which suggests the origin of these may be physical. Inspecting the explosive behaviour in detail, we find that a very localised intense process occurs at the current sheet of the magnetic island at the LMRS. In both local simulations, and simulations with a narrower global domain, runaway is not seen, indicating that this process is somehow sensitive to boundary conditions or spatial background nonuniformity.



\begin{figure}[t]
\centering
\includegraphics[width=\columnwidth] {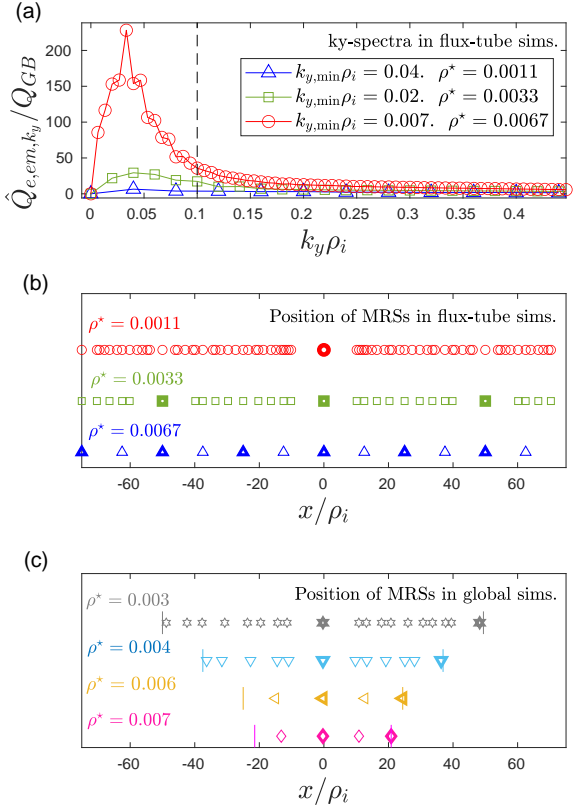}
\caption{(a) $k_y-$spectra of electron electromagnetic heat flux for the flux-tube simulations. Radial position of MRSs for all $k_y\rho_i\leq 0.1$ for (b) flux-tube and (c) global simulations with $L_x=0.3a$. Thicker markers denote LMRSs. Vertical lines denote radial box boundaries.}
\label{Figspectra}
\end{figure}

To isolate the cause of this system-size dependence, \emph{i.e.} the increase in heat flux with increasing system-size, we use flux-tube simulations.
Note that by setting $N_{pol}=1$ and by choosing $k_{y,\min}\rho_i= n_{\min}q_{0}(a/x_0)\rho^\star$ corresponding to the fundamental mode ($n_{\min}=1$) of the tokamak, one can simulate the full flux-surface of interest in the tokamak using flux-tube. This helps to capture the correct parallel connection length along the magnetic field lines when the LMRSs correspond to an integer rational surface. Furthermore, the radial distance between rational surfaces is also correctly captured. See Figs.~\ref{Figspectra}(b) and (c) for comparing the distance between (L)MRSs in flux-tube and global simulations respectively. Given that self-interaction is essentially `modes biting their tails' at rational surfaces, the flux-tube framework offers the possibility to accurately capture the effect of self-interaction, while neglecting other finite $\rho^*$ effects such as profile shearing (see Refs.~\cite{Justin2020,AjayCJ2020} for more details). 
Given that the separation distance between the LMRSs is $1/(\hat{s}k_{y,\min})$ in flux-tube simulations, we can study this particular system-size effect through a scan in $k_{y,\min}\rho_i$. Both $L_x/\rho_i$  and $k_{y,\max}\rho_i$ are kept fixed in this scan.

As $k_{y,\min}$ is decreased, the radial density of regions with flattened electron temperature (see Fig.~\ref{Fig3}(a)), and hence weaker linear drive, at low-order MRSs decreases. Concurrently, flux increases, as shown by the blue asterisks in Fig.~\ref{Fig5}.
That is, the temperature flattening mechanism becomes less effective in large systems. For the $k_{y,\min}\rho_i=0.02$ case, the $k_y\rho_i=0.04$ mode contributing most to the flux has six MRSs, three of which at LMRSs experience $\sim\!\! 70\%$ flattening, and the other three at second-order MRSs experience $\sim\!\! 10\%$ flattening. Whereas for $k_{y,\min}\rho_i=0.04$, the $k_y\rho_i=0.04$ mode sees a $\sim\!\! 70\%$ temperature flattening at every MRS, so mode stabilization is much more effective. When the electron temperature flattenings are eliminated, there is still some non-gyro-Bohm scaling (black markers in Fig.~\ref{Fig5}), but this is less consistent.

We suggest a crude model to understand the increase in flux with increasing system-size that persists in the local limit. In the turbulent steady state, when the electron heat flux $Q_e$ becomes radially constant, one defines the pointwise diffusivity via $ \chi_e \equiv Q_e / \, (dT_e/dx)$ and the radial average
\begin{equation}
\left\langle\frac{dT_e}{dx}\right\rangle_x
= \left\langle\frac{Q_e}{\chi_e}\right\rangle_x
= Q_e\left\langle\frac{1}{\chi_e}\right\rangle_x.
\end{equation}
so $\langle 1/\chi_e\rangle_x^{-1}$ is the effective average diffusivity. 
This analysis also holds in the source-free region of a tokamak or global simulation. In the local limit, boundary conditions impose zero average temperature fluctuation, thus $Q_e = {dT_{e,0}}/{dx} \langle 1/\chi_e\rangle_x^{-1}$.

Microtearing modes are now modelled to lead to regions of high diffusivity near each MRS, which reinforce at LMRSs, resulting in the temperature-gradient corrugations seen in simulations.

The set of `relevant' MRSs, i.e. associated with all the modes contributing significant flux, becomes more radially dense with decreasing $k_{y,\min}$, because the number of toroidal modes increases, and each mode has associated rationals separated by $1/\hat{s} k_y$. This is illustrated in Fig~\ref{Figspectra}. In Fig~\ref{Figspectra}(a), $k_y-$spectra of the electron electromagnetic heat flux for the flux-tube simulations are plotted for the three considered values of $k_{y,\min}$. $\hat{Q}_{e,em,k_y}$ is defined such that the total flux $Q_{e,em}=\sum_{k_y}\hat{Q}_{e,em,k_y} {\rm d}k_y$, where ${\rm d}k_y=k_{y,\min}\rho_i$. In Fig~\ref{Figspectra}(b), the positions of all MRSs associated with $k_y\rho_i\leq 0.1$ (\emph{i.e.} contributing significant flux) are marked for each of the three simulations, clearly indicating how the MRSs become radially dense with decreasing $k_{y,\min}$. The opposite is true for LMRSs (the MRSs common to all $k_y$s and separated by $1/\hat{s}k_{y,\min}$), denoted by thicker markers in Fig~\ref{Figspectra}(b), which become more spaced apart with decreasing $k_{y,\min}$. For comparison, the corresponding plot for global simulations with $L_x=0.3a$ is shown in Fig~\ref{Figspectra}(c).

As the harmonic mean of diffusivity sets flux levels, concentrating the diffusivity at widely-spaced MRSs (large $k_{y,\min}$, small system-size) leads to lower flux than distributing it more evenly at a larger number of closely-spaced MRSs (low $k_{y,\min}$, large system-size). In an extreme limit, microtearing creates infinite local diffusivity and completely flattens gradients near each MRS, but elsewhere the diffusivity is a small constant $\chi_b$. The effective average diffusivity, crudely assuming no overlap between flattened regions, is $\chi_b/ (1 - W N)$, where $W$ is the proportion of the radius flattened by each toroidal mode and $N$ is the number of toroidal modes. This leads to a scaling $Q_e \propto 1/(1 - w/\rho^\star)$ where $w$ a dimensionless small parameter that depends on parameters other than $\rho^\star$; note that the flux has a singularity at small enough $\rho^\star$, where the flux explodes. This might be related to the increase in flux with system size seen (Fig.~\ref{Fig5}) in our MT simulations. 
This is also analogous to the avalanche transport arising when transport windows caused by fast-particle-driven modes overlap across much of the tokamak~\cite{Fredrickson_2006}. 

The width $W$ of the high-transport region was assumed fixed in this simple picture, but actually may increase with higher flux, and the larger overlap may cause a runaway situation; this may be tied to failure to reach saturation in certain microtearing simulations. 

Apart from system-size scaling, another possible consequence of electron temperature flattenings and magnetic islands at low-order rational surfaces is the potential to seed the growth of NTMs \cite{Buttery2000}. The possibility for microturbulence to excite NTMs via nonlinear coupling has been demonstrated in the past \cite{Muraglia2011}. Furthermore, to experimentally verify our results, one may measure the electron temperature and look for flattenings near rational surfaces, similar to previous investigations of ITG turbulence \cite{Waltz2006}. One may also be able to measure low toroidal mode number magnetic perturbations associated with the microtearing islands in external magnetic coils; for instance, the radial magnetic perturbation associated with the islands is $(\delta B_x/B_0)/(\rho_i/R)\simeq 0.14$ in the standard nonlinear simulation.

\section{Conclusions}
In conclusion, the fast motion of electrons across the magnetic islands at the LMRSs short-circuit the electron temperature, resulting in local electron temperature flattening, which then decreases the local linear drive of microtearing modes and allows lower saturated transport levels than when this process is artificially suppressed. The spacing and width of the low-confinement regions near low-order rationals are crucial, and this provides a pathway to understand microtearing saturation (or lack thereof); one direct consequence is that microtearing turbulent transport and its study are more important in larger future devices than previously thought. 

Note that microtearing saturation mechanisms that depend on zonal corrugations of other quantities, such as zonal fields or flows at rationals, would also be expected to show the same non-gyro-Bohm dependence, due to the same density-of-rationals argument.

We note that there are various important unanswered questions about size-scaling and global effects for microtearing transport. For example, the transport level is also found to be sensitive to the nature (global versus local) and radial extent of the simulation domain. This is perhaps not surprising given that the field perturbations of microtearing modes are radially extended, even though they cause localised transport. Also, we do not know how the runaway process occurs, except that it is the result of a localised intense event in the current sheet around a magnetic island. Only some of the simulations here seem subject to this process, and this also provides some clues.


\section*{Acknowledgements}
We acknowledge Tobias Goerler and Alessandro Di Siena for helping us set up global simulations. We acknowledge the CINECA award under the ISCRA initiative, for the availability of high performance computing resources and support.

\section*{References}
\bibliography{MTM_paper_3} 

\begin{thebibliography}{10}

\bibitem{Hazeltine1975}
R.~D. Hazeltine, D.~Dobrott, and T.~S. Wang.
\newblock Kinetic theory of tearing instability.
\newblock {\em The Phys. Fluids}, 18(12):1778--1786, 1975.

\bibitem{Drake1980}
J.~F. Drake, N.~T. Gladd, C.~S. Liu, and C.~L. Chang.
\newblock Microtearing modes and anomalous transport in tokamaks.
\newblock {\em Phys. Rev. Lett.}, 44:994--997, 1980.

\bibitem{Connor1990}
J.~W. Connor, S.~C. Cowley, and R.~J. Hastie.
\newblock Micro-tearing stability in tokamaks.
\newblock {\em Plasma Phys. Controlled Fusion}, 32(10):799--817, 1990.

\bibitem{Gryaznevich1998}
M.~Gryaznevich, R.~Akers, P.~G. Carolan, N.~J. Conway, D.~Gates, A.~R. Field,
  T.~C. Hender, I.~Jenkins, R.~Martin, M.~P.~S. Nightingale, C.~Ribeiro, D.~C.
  Robinson, A.~Sykes, M.~Tournianski, M.~Valovic\ifmmode~\breve{}\else
  \u{}\fi{}, and M.~J. Walsh.
\newblock Achievement of record $\mathit{\ensuremath{\beta}}$ in the start
  spherical tokamak.
\newblock {\em Phys. Rev. Lett.}, 80:3972--3975, May 1998.

\bibitem{Applegate2004}
D.~J. Applegate, C.~M. Roach, S.~C. Cowley, W.~D. Dorland, N.~Joiner, R.~J.
  Akers, N.~J. Conway, A.~R. Field, A.~Patel, M.~Valovic, and M.~J. Walsh.
\newblock Microstability in a {MAST-like} high confinement mode spherical
  tokamak equilibrium.
\newblock {\em Phys. Plasmas}, 11(11):5085--5094, 2004.

\bibitem{Guttenfelder2011}
W.~Guttenfelder, J.~Candy, S.~M. Kaye, W.~M. Nevins, E.~Wang, R.~E. Bell, G.~W.
  Hammett, B.~P. LeBlanc, D.~R. Mikkelsen, and H.~Yuh.
\newblock Electromagnetic transport from microtearing mode turbulence.
\newblock {\em Phys. Rev. Lett.}, 106:155004, 2011.

\bibitem{Patel2021}
B.S. Patel, D.~Dickinson, C.M. Roach, and H.R. Wilson.
\newblock Linear gyrokinetic stability of a high beta non-inductive spherical
  tokamak.
\newblock {\em Nucl. Fusion}, 62(1):016009, 2021.

\bibitem{Wilson2020}
H.~Wilson, I.~Chapman, T.~Denton, W.~Morris, B.~Patel, G.~Voss, C.~Waldon, and
  the STEP~Team.
\newblock {STEP}-on the pathway to fusion commercialization.
\newblock In {\em Commercialising Fusion Energy}, 2053-2563, pages 8--1 to
  8--18. IOP Publishing, 2020.

\bibitem{Hatch2021}
D.~R. Hatch, M.~Kotschenreuther, S.~M. Mahajan, M.~J. Pueschel, C.~Michoski,
  G.~Merlo, E.~Hassan, A.~R. Field, L.~Frassinetti, C.~Giroud, J.C. Hillesheim,
  C.F. Maggi, C.~Perez von Thun, C.M. Roach, S.~Saarelma, D.~Jarema, F.~Jenko,
  and JET Contributors.
\newblock Microtearing modes as the source of magnetic fluctuations in the
  {JET} pedestal.
\newblock {\em Nucl. Fusion}, 61(3):036015, 2021.

\bibitem{Drake1981}
J.~F. Drake and A.~B. Hassam.
\newblock Collisional drift waves in a plasma with electron temperature
  inhomogeneity.
\newblock {\em Phys. Fluids}, 24(7):1262--1269, 1981.

\bibitem{Carmody2013}
D.~Carmody, M.~J. Pueschel, and P.~W. Terry.
\newblock Gyrokinetic studies of microinstabilities in the reversed field
  pinch.
\newblock {\em Phys. Plasmas}, 20(5):052110, 2013.

\bibitem{Hamed2018_2}
M.~Hamed, M.~Muraglia, Y.~Camenen, X.~Garbet, and O.~Agullo.
\newblock Impact of electric potential and magnetic drift on microtearing modes
  stability.
\newblock {\em Phys. Plasmas}, 26(9):092506, 2019.

\bibitem{Drake1977}
J.~F. Drake and Y.~C. Lee.
\newblock Kinetic theory of tearing instabilities.
\newblock {\em Phys. Fluids}, 20(8):1341--1353, 1977.

\bibitem{Gladd1980}
N.~T. Gladd, J.~F. Drake, C.~L. Chang, and C.~S. Liu.
\newblock Electron temperature gradient driven microtearing mode.
\newblock {\em Phys. Fluids}, 23(6):1182--1192, 1980.

\bibitem{Predebon2013}
I.~Predebon and F.~Sattin.
\newblock On the linear stability of collisionless microtearing modes.
\newblock {\em Phys. Plasmas}, 20(4):040701, 2013.

\bibitem{Swamy2015}
A.~K. Swamy, R.~Ganesh, S.~Brunner, J.~Vaclavik, and L.~Villard.
\newblock Collisionless microtearing modes in hot tokamaks: Effect of trapped
  electrons.
\newblock {\em Phys. Plasmas}, 22(7):072512, 2015.

\bibitem{Doerk2011}
H.~Doerk, F.~Jenko, M.~J. Pueschel, and D.~R. Hatch.
\newblock Gyrokinetic microtearing turbulence.
\newblock {\em Phys. Rev. Lett.}, 106:155003, 2011.

\bibitem{Maeyama2017_2}
S.~Maeyama, T.-H. Watanabe, and A.~Ishizawa.
\newblock Suppression of ion-scale microtearing modes by electron-scale
  turbulence via cross-scale nonlinear interactions in tokamak plasmas.
\newblock {\em Phys. Rev. Lett.}, 119:195002, 2017.

\bibitem{Pueschel2020}
M.~J. Pueschel, D.~R. Hatch, M.~Kotschenreuther, A.~Ishizawa, and G.~Merlo.
\newblock Multi-scale interactions of microtearing turbulence in the tokamak
  pedestal.
\newblock {\em Nucl. Fusion}, 60(12):124005, 2020.

\bibitem{DoerkPhD}
H.~Doerk.
\newblock {\em Gyrokinetic simulation of microtearing turbulence}.
\newblock PhD thesis, Universit{\"a}t Ulm, 2012.

\bibitem{Xie2020}
T.~Xie, M.~J. Pueschel, and D.~R. Hatch.
\newblock Quasilinear modeling of heat flux from microtearing turbulence.
\newblock {\em Phys. Plasmas}, 27(8):082306, 2020.

\bibitem{Hardman2022}
M.~R. Hardman, F.~I. Parra, C.~Chong, T.~Adkins, M.~S. Anastopoulos-Tzanis,
  M.~Barnes, D.~Dickinson, J.~F. Parisi, and H.~Wilson.
\newblock Extended electron tails in electrostatic microinstabilities and the
  nonadiabatic response of passing electrons.
\newblock {\em Plasma Phys. Controlled Fusion}, 64(5):055004, 2022.

\bibitem{GENE1}
F.~Jenko, W.~Dorland, M.~Kotschenreuther, and B.~N. Rogers.
\newblock Electron temperature gradient driven turbulence.
\newblock {\em Phys. Plasmas}, 7(5):1904--1910, 2000.

\bibitem{GENE2}
T.~G{\"o}rler, X.~Lapillonne, S.~Brunner, T.~Dannert, F.~Jenko, F.~Merz, and
  D.~Told.
\newblock The global version of the gyrokinetic turbulence code gene.
\newblock {\em J. Comput. Phys.}, 230(18):7053 -- 7071, 2011.

\bibitem{Beer1995}
M.~A. Beer, S.~C. Cowley, and G.~W. Hammett.
\newblock Field-aligned coordinates for nonlinear simulations of tokamak
  turbulence.
\newblock {\em Phys. Plasmas}, 2(7):2687--2700, 1995.

\bibitem{Scott1998}
B.~Scott.
\newblock Global consistency for thin flux tube treatments of toroidal
  geometry.
\newblock {\em Phys. Plasmas}, 5(6):2334--2339, 1998.

\bibitem{GoerlerPhD}
T.~{G{\"o}rler}.
\newblock {\em Multiscale effects in plasma microturbulence}.
\newblock PhD thesis, Universit{\"a}t Ulm, 2009.

\bibitem{McMillan2010}
B.~F. McMillan, X.~Lapillonne, S.~Brunner, L.~Villard, S.~Jolliet, A.~Bottino,
  T.~G\"orler, and F.~Jenko.
\newblock System size effects on gyrokinetic turbulence.
\newblock {\em Phys. Rev. Lett.}, 105:155001, Oct 2010.

\bibitem{Lapillonne2009}
X.~Lapillonne, S.~Brunner, T.~Dannert, S.~Jolliet, A.~Marinoni, L.~Villard,
  T.~G{\"o}rler, F.~Jenko, and F.~Merz.
\newblock Clarifications to the limitations of the s-alpha equilibrium model
  for gyrokinetic computations of turbulence.
\newblock {\em Phys. Plasmas}, 16(3):032308, 2009.

\bibitem{Lapillonne2011_2}
X.~Lapillonne, B.~F. McMillan, T.~Görler, S.~Brunner, T.~Dannert, F.~Jenko,
  F.~Merz, and L.~Villard.
\newblock Nonlinear quasisteady state benchmark of global gyrokinetic codes.
\newblock {\em Physics of Plasmas}, 17(11):112321, 2010.

\bibitem{Nevins2011}
W.~M. Nevins, E.~Wang, and J.~Candy.
\newblock Magnetic stochasticity in gyrokinetic simulations of plasma
  microturbulence.
\newblock {\em Phys. Rev. Lett.}, 106:065003, 2011.

\bibitem{Pueschel2013_2}
M.~J. Pueschel, D.~R. Hatch, T.~G\"orler, W.~M. Nevins, F.~Jenko, P.~W. Terry,
  and D.~Told.
\newblock Properties of high-beta microturbulence and the non-zonal transition.
\newblock {\em Phys. Plasmas}, 20(10):102301, 2013.

\bibitem{Justin2020}
J.~Ball, S.~Brunner, and {Ajay C. J.}
\newblock Eliminating turbulent self-interaction through the parallel boundary
  condition in local gyrokinetic simulations.
\newblock {\em J. Plasma Phys.}, 86(2):905860207, 2020.

\bibitem{AjayCJ2020}
{Ajay C. J.}, S.~Brunner, B.~McMillan, J.~Ball, J.~Dominski, and G.~Merlo.
\newblock How eigenmode self-interaction affects zonal flows and convergence of
  tokamak core turbulence with toroidal system size.
\newblock {\em J. Plasma Phys.}, 86(5):905860504, 2020.

\bibitem{AjayCJ2021}
{Ajay C. J.}, S.~Brunner, and J.~Ball.
\newblock Effect of collisions on non-adiabatic electron dynamics in
  {ITG-}driven microturbulence.
\newblock {\em Phys. Plasmas}, 28(9):092303, 2021.

\bibitem{Hallatscheck2005}
K.~Hallatschek and W.~Dorland.
\newblock Giant electron tails and passing electron pinch effects in
  tokamak-core turbulence.
\newblock {\em Phys. Rev. Lett.}, 95:055002, 2005.

\bibitem{Pueschel2013_3}
M.~J. Pueschel, T.~G\"orler, F.~Jenko, D.~R. Hatch, and A.~J. Cianciara.
\newblock On secondary and tertiary instability in electromagnetic plasma
  microturbulence.
\newblock {\em Phys. Plasmas}, 20(10):102308, 2013.

\bibitem{Ishizawa2014}
A.~Ishizawa, T.-H. Watanabe, H.~Sugama, S.~Maeyama, and N.~Nakajima.
\newblock Electromagnetic gyrokinetic turbulence in finite-beta helical
  plasmas.
\newblock {\em Phys. Plasmas}, 21(5):055905, 2014.

\bibitem{Fredrickson_2006}
E.~D Fredrickson, N.~N Gorelenkov, R.~E Bell, J.~E Menard, A.~L Roquemore,
  S.~Kubota, N.~A Crocker, and W.~Peebles.
\newblock Fast ion loss in a `sea-of-{TAE}'.
\newblock {\em Nucl. Fusion}, 46(10):S926--S932, 2006.

\bibitem{Buttery2000}
R.~J. Buttery, S.~G{\"u}nter, G.~Giruzzi, T.~C. Hender, D.~Howell, G.~Huysmans,
  R.~J. La~Haye, M.~Maraschek, H.~Reimerdes, O.~Sauter, C.~D. Warrick, H.~R.
  Wilson, and H.~Zohm.
\newblock Neoclassical tearing modes.
\newblock {\em Plasma Phys. Controlled Fusion}, 42(12B):B61--B73, 2000.

\bibitem{Muraglia2011}
M.~Muraglia, O.~Agullo, S.~Benkadda, M.~Yagi, X.~Garbet, and A.~Sen.
\newblock Generation and amplification of magnetic islands by drift interchange
  turbulence.
\newblock {\em Phys. Rev. Lett.}, 107:095003, 2011.

\bibitem{Waltz2006}
R.~E. Waltz, M.~E. Austin, K.~H. Burrell, and J.~Candy.
\newblock Gyrokinetic simulations of off-axis minimum-q profile corrugations.
\newblock {\em Phys. Plasmas}, 13(5):052301, 2006.

\end{thebibliography}
\bibliographystyle{unsrt}

\end{document}